\documentclass[12pt,preprint]{aastex}

\shorttitle{3-Dimensional Structure of HH 32}
\shortauthors{Beck et al.}

\begin{document}
\title{The 3-Dimensional Structure of HH 32 from GMOS IFU Spectroscopy\altaffilmark{1}}
\author{Tracy L. Beck,\altaffilmark{2} A. Riera,\altaffilmark{3,4,5} A. C. Raga,\altaffilmark{6} C. Aspin,\altaffilmark{2}}

\email{tbeck@gemini.edu, angels.riera@upc.es, raga@nuclecu.unam.mx, caa@gemini.edu }

\altaffiltext{1}{Based on observations obtained at the Gemini Observatory, which is operated by the Association of Universities for Research in Astronomy, Inc., under a cooperative agreement with the NSF on behalf of the Gemini partnership: the National Science Foundation (United States), the Particle Physics and Astronomy Research Council (United Kingdom), the National Research Council (Canada), CONICYT (Chile), the Australian Research Council (Australia), CNPq (Brazil), and CONICET (Argentina).}

\altaffiltext{2}{Gemini Observatory, Northern Operations, 670 N. A'ohoku Pl., Hilo, HI, 96720}
\altaffiltext{3}{Departament de F\'\i sica i Enginyeria Nuclear, Universitat Polit\'ecnica de Catalunya, Av. V\'\i ctor 
Balaguers/n E-08800 Vilanova i La Geltr\'u, Spain }
\altaffiltext{4}{Departament d'Astronomia i Meteorologia, Universitat de Barcelona, Av. Diagonal 647, E-08028 Barcelona, Spain}
\altaffiltext{5}{On sabbatical leave at the Instituto de Ciencias Nucleares, UNAM}
\altaffiltext{6}{Instituto de Ciencias Nucleares, UNAM, Ap. 70-543, 04510 D.F., M\'exico}

\begin{abstract}
We present new high resolution spectroscopic observations of the Herbig-Haro object HH~32 from System Verification observations made with the GMOS IFU at Gemini North Observatory.  The 3D spectral data covers a $8''.7\times 5''.85$ spatial field and 4820 - 7040 \AA\ spectral region centered on the HH~32 A knot complex.  We show the position-dependent line profiles and radial velocity channel maps of the H$\alpha$ line, as well as line ratio velocity channel maps of [O~III]~5007/H$\alpha$, [O~I]~6300/H$\alpha$, [N~II]~6583/H$\alpha$, [S~II]~(6716+6730)/H$\alpha$ and [S~II]~6716/6730.  We find that the line emission and the line ratios vary significantly on spatial scales of $\sim$1$''$ and over velocities of $\sim$50 km/s.   A ``3/2-D'' bow shock model is qualitatively successful at reproducing the general features of the radial velocity channel maps, but it does not show the same complexity as the data and it fails to reproduce the line ratios in our high spatial resolution maps.  The observations of HH~32 A show two or three superimposed bow shocks with separations of $\sim 3''$, which we interpret as evidence of a line of sight superposition of two or three working surfaces located along the redshifted body of the HH~32 outflow.
\end{abstract}



\section{Introduction}

HH~32 is one of the brightest sources in the original catalog of Herbig-Haro objects (Herbig 1974).  Spectroscopic studies have shown that HH~32 has a high excitation spectrum with significant emission from H, He and a wide range of metals in the 3700 to 10800~\AA\ region (Dopita 1978; Brugel, B\"ohm \& Mannery 1981; Raga, B\"ohm \& Cant\'o 1996).  HH 32 also has an unusually large extinction (E(B-V)$\approx 0.7$; Brugel, B\"ohm \& Mannery 1981), hence it is surprising that this object was detected in the ultraviolet with IUE (B\"ohm \& B\"ohm-Vitense 1984; Lee et al. 1988; Moro-Mart\'\i n et al. 1996).  The UV detection results from of the strong emission in the high exitation [C~III]~1909~\AA\ and [C~IV]~1550~\AA\ ultraviolet lines.   Interestingly, this high excitation object also shows H$_2$ {\it v} = 1-0 S(1) emission in the infrared which is excited in shocks as the flow encounters ambient cloud material. (Zealey et al. 1986; Zinnecker et al. 1989; Davis, Eisl\"offel \& Smith 1996).

The HH~32 outflow is clearly associated with the T~Tauri star AS~353A, with condensations A, B and D at $\sim 20''$ to the west, and condensation C at $\sim 10''$ to the east of the star (see Figure 1 from Curiel et al. 1997).   AS~353A has a rich emission line spectrum (Eisl\"offel, Solf \& B\"ohm 1990), but shows only two forbidden emission lines: [O~I]~5577, 6300~\AA.  High resolution optical spectroscopy of the HH condensations has shown that HH~32 A, B and D have very broad, redshifted line profiles, with full widths of $\sim 400$~km~s$^{-1}$ (Herbig \& Jones 1983; Hartigan, Mundt \& Stocke 1986; Solf, B\"ohm \& Raga 1986; Hartigan, Raymond \& Hartmann 1987).  The faint knot C to the east of the exciting source is blueshifted, indicating that the HH 32 emission knots are part of a bipolar outflow from AS 353A.    The very large line widths, the shape of the line profiles, and the position-velocity diagrams have been interpreted successfully in terms of ``3/2-D'' bow shock models (Solf et al. 1986; Raga, B\"ohm \& Solf 1986; Raga \& B\"ohm 1986; Hartigan et al. 1987). In such models, the bow shock is modeled with a parametrized shape (i.e., Hartigan et al. 1987) and the emission is computed by assuming that the post-bow shock recombination region resembles the structure of a 1D stationary shock.   Comparisons with the models show a $v_{bs}\sim 300-350$~km~s$^{-1}$ velocity for the bow shock (Solf et al. 1986; Hartigan et al. 1987). 

By comparing the proper motions with the radial velocities we know that HH~32A is moving away from the observer at a $\phi\approx 70^\circ$ angle with respect to the plane of the sky (Herbig \& Jones 1983).  This angle is consistent with the orientation necessary to model the emission line profiles (Solf et al. 1986; Hartigan et al. 1987).   The detailed morphological structure and the proper motions of the sub-condensations of HH~32A described by Curiel et al. (1997) have been interpreted in terms of 3/2-D bow shock models to obtain the same $\phi=70^\circ$ and $v_{bs}\sim 300$~km~s$^{-1}$ values that are deduced from the emission line profiles (Raga et al. 1997).

In this paper we present and discuss new spectroscopic observations with 2D spatial coverage of a field centered on the HH~32A condensation, which give new information on the kinematics and structure of the emission knots.  The previous spatially resolved spectroscopy of Solf et al. (1986) covered all of HH~32 with several different long-slit positions centered on AS~353A, but did not have a comparable angular resolution nor as high a signal-to-noise ratio as the spectra we present here.  Also, the HST images of Curiel et al. (1997) show the high resolution spatial structure of HH~32, but lack any kinematic information.  In $\S$2, we present the observations and discuss briefly the method of data reduction.  We show the H$\alpha$ line profiles and velocity channel maps obtained from the spectra, and the velocity channel maps of different line ratios in $\S$3 and $\S$4.  In $\S$5 we interpret the observations in terms of a 3/2-D bow shock model and discuss its successes and failures in modeling the data.  Finally, we present our conclusions and a discussion of the implications in $\S$6.

\section{The observations}

The data for this project were obtained on 2002 Aug 03 with the Gemini Multi-Object Spectrograph (GMOS; Hook et al. 2002) Integral Field Unit (IFU; Allington-Smith et al. 2002) at the Gemini North Fredrick C. Gillett Telescope.  These data were taken as System Verification for dithered observations on an emission line source, and are available to the public at: http://www.gemini.edu/sciops/instruments/gmos/SVobs/gmosSVobs73.html under Gemini program ID GN-2002B-SV-73.   The observations were made in cloudy conditions (0.5 to 1 magnitude of extinction) with $\sim$0.$''$5 average seeing.  The source star of the HH 32 outflow, AS 353A, was used to provide guiding and tip-tilt corrections using the GMOS on-instrument wavefront sensor (OIWFS).

These data were obtained with the IFU in 1-slit mode, which provides a spatial field of $3\farcs5$ x 5$''$ for each resulting science datacube.  With this observing configuration, the GMOS IFU is comprised of 750 fibers; each spans a $0\farcs2$ hexagonal region on the sky.  500 fibers make up the  $3\farcs5$ x 5$''$ science field of view, and 250 fibers make up a smaller, dedicated sky field which is fixed at a 1$'$ distance from the science position (Allington-Smith et al. 2002).  A total of 12 900 second frames were obtained; two steps through a 3 x 2 dither pattern resulting in total integration time of 30 minutes at each of the 6 dither positions.   We used the R831 grating in GMOS, which has a 0.034 nm/pixel scale and results in $\approx 15$~km~s$^{-1}$ spectral resolution.  We were able to obtain spectra over the 4920 - 7040 \AA\ region, which includes the [O~III]~4969/5007, [O~I]~6300/63, [N~II]~6548/83, H$\alpha$, and [S~II]~6716/31 lines.  

We used the GMOS IFU scripts available in the Gemini IRAF package for reduction of the data.  The {\sc gfreduce} task was used to prepare the frames for reduction, subtract off the bias and overscan levels, reject cosmic rays by calling the {\sc gscrrej} routine, and call the additional reduction routines {\sc gfextract} and {\sc gfskysub}.  {\sc gfextract} traces and extracts each IFU fiber to 1-D spectral output in the flat field image, applies this trace to the science data, extracts the science spectra using a $\pm$2.5 pixel aperture on each side of the traces, and applies the flat field correction.  {\sc gfskysub} subtracts off a sky spectrum from the science data using averaged spectra from the dedicated sky field inherent in the IFU.  The {\sc gswavelength} task then establishes the wavelength calibration for the IFU arc lamp images, which is thus applied to the science frames using the {\sc gftransform} routine.  These 2-Dimensional data images were reformatted into a 3-D datacube format (x, y, $\lambda$) and resampled to square pixels with a 0.$''$05 spatial resolution using the {\sc gfcube} routine.  The dithered data were mosaiced together to form a larger cube in the spatial dimension by manually offsetting and coadding the individual cubes using IDL.   The final datacube has a spatial extent of 8$''$.7 by 5.$''$85.  At the forementioned webpage we have made available the raw data, the final IRAF datacubes, the mosaiced cube constructed in IDL, an IRAF cl-reduction script, and a README file that describes our data reduction process.

In Figure 1 we present a 30 second exposure r-band image of the HH 32 system that was obtained for setup on the IFU field; the HH 32 A1, A2, B and D condensations are labeled (using the notation of Curiel et al. 1997).  Overplotted on the image is the position and orientation in the spatial dimension of the final mosaiced datacube, and a vector indicating the direction to the outflow source, AS 353A.   In Figure 2 we show a spectrum of the HH 32 A2 knot averaged over the inner $1\farcs5$ in the spatial dimension; the [O~I]~6300/63, [N~II]~6548/83, H$\alpha$ and [S~II]~6716/31 lines are identified.  All of the lines show a pronounced double-peaked structure, consistent with previous observations (Hartigan, Mundt \& Stocke 1986).

\section{H$\alpha$ line profiles and channel maps}

Figure 3 shows a contour map of the H$\alpha$ emission constructed by summing the line over its wavelength extent in the final mosaiced datacube.  Overplotted in Figure~3 are the H$\alpha$ line profiles obtained by integrating the emission over boxes of $10\times 10$ pixels (corresponding to $0\farcs5 \times 0\farcs5$).  Many of the observed line profiles are double peaked, with a stronger peak at a $v_r\approx 90$~km~s$^{-1}$ heliocentric radial velocity and a weaker high velocity peak at $v_r\approx 240\to 276$~km~s$^{-1}$.  Several of the line profiles in the emission joining the maxima of the A1 and A2 condensations show a dominant high velocity peak, and a few of the profiles in the periphery of A1 show three components.  In many of the line profiles along the edges of the observed field, only the low velocity component is visible.

In Figure 4 we present the H$\alpha$ radial velocity channel maps derived from the IFU observations.  The H$\alpha$ emission is detected from heliocentric radial velocities ranging from $v_r\approx -20$ to $+400$~km~s$^{-1}$.  The surface brightness of the condensations grows with increasing $v_r$, reaching a maximum emission level in the $+72$~km~s$^{-1}$ map.  It then decreases, reaching a minimum at $+165$~km~s$^{-1}$, and increases again reaching a second maximum at $v_r=+241$~km~s$^{-1}$.  For the larger radial velocities, the surface brightness decreases monotonically.

At $v_r=10$~km~s$^{-1}$, there are two condensations with a separation of $\approx 2\farcs40$.  The western condensation (A2) has a structure of two superimposed arcs and the eastern condensation (A1) also has an arc-like structure.   For both condensations, this basic morphology is preserved to velocities of $v_r\approx 100$~km~s$^{-1}$.  We find a pronounced arc of emission that connects the A1 and A2 condensations in the $v_r \approx 25$~km~s$^{-1}$ to $v_r \approx 87$~km~s$^{-1}$ maps.   A1 and A2 no longer have their arc-like shapes and have a more compact morphology for $v_r> 118$~km~s$^{-1}$.   They approach each other in the spatial direction for increasing values of $v_r$, and have a separation of only $\approx 1\farcs15$ in the $v_r=304$~km~s$^{-1}$ map.  Condensation A1 dominates the total intensity for $v_r> 300$~km~s$^{-1}$, and again develops a compact, arc-like shape for higher values of $v_r$.  

Although not shown, the channel maps for the [O~I], [N~II], and [S~II] lines show qualitatively similar morphologies as those for H$\alpha$ .  However, the channel maps for the [O~III] line show significant differences.  The surface brightness of the two condensations in [O III] grows with increasing radial velocity, reaching a maximum level at  +168 km s$^{-1}$, and then the surface brightness decreases for larger velocities. No second maximum is detected.  At the lowest radial velocities, subcondensation A2 shows an elongated structure that, despite its low signal-to-noise, resembles the arc-shaped structure also visible in the H$\alpha$ maps. At higher radial velocities, subcondensation A1 has a more compact morphology.  Subcondensation A1 is detected at $v_r$ from +26 to +350  km s$^{-1}$, and it dominates the total intensity for $v_r$ $>$ +188 km s$^{-1}$.  Subcondensation A1 shows a compact morphology at these radial velocities; no arc-like structure is detected in this subcondensation in the [O III] emission.  As reported for the H$\alpha$ maps, both subcondenstions approach each other in the spatial direction as the radial velocity increases.

\section{The line ratios}

The [S~II]~6716/6731 ratio depends only weakly on temperature, so it is a direct diagnostic of the electron density of the emitting plasma.  Unfortunately, the other ratios between the observed lines do not have such a direct interpretation.  However, ratios such as [S~II]~(6716+6731)/H$\alpha$ and [O~III]~5007/H$\alpha$ do show clear trends as a function of shock velocity in models of HH jets (see, e.~g., Hartigan et al. 1987), and can therefore be used to estimate the velocities of the shocks which dominate the emission along a given line of sight. Our spectra are not flux calibrated, so we are not able to give direct estimates of the values of the shock velocities.  Therefore, our results can only be used to obtain a qualitative picture of the variations of the shock properties across the emitting region of HH~32.

From the position-velocity (PV) cubes of the [O~I]~6300, [N~II]~6583, [S~II]~6716, 6731, [O~III]~5007 and the H$\alpha$ lines, we have generated seven channel maps centered at heliocentric radial velocities of $v_r=-5$, 48, 110, 172, 234, 296 and 358~km~s$^{-1}$ (each map has a velocity width of 47~km~s$^{-1}$). We have binned the maps spatially over $5\times 5$ pixels of the original spectrum to decrease the effective pixel size to $0\farcs25$.   Figure~5 shows the [O~I]~6300 and [N~II]~6583 channel maps divided by the corresponding H$\alpha$ maps, Figure~6 shows the [S~II]~(6716+6731)/H$\alpha$ and the [S~II]~6716/6731 line ratio channel maps and Figure 7 shows the [O III] 5007 \AA\ channel maps divided by the corresponding H$\alpha$ channel maps.  The overall spatial size of the line ratio maps is the same as the H$\alpha$ channel maps, but the pixels have been binned by 5 in the x and y directions.  The line ratio maps appear to cover a wider area than the H$\alpha$ emission maps because there is non-negligible flux in the lines at distances spatially extended from the condensations seen in H$\alpha$.  As a function of radial velocity, the behavior of the line ratio maps can be described as:

\begin{itemize}
\item [N~II]~6583/H$\alpha$~: at the position of knot A2, this line ratio has low values of $\approx 0.4$ at low radial velocities, and the ratio grows with increasing $v_r$.  At the position of knot A1, the maximum values of $\approx 0.7$ for the line ratio occur at $v_r=48$~km~s$^{-1}$, and the ratios decrease for increasing $v_r$, reaching values of $\approx 0.4$ for $v_r=296$~km~s$^{-1}$.  Interestingly, the region with the highest line ratio lies to the South of knot A1, with values of $\approx 1$ for $v_r=48$~km~s$^{-1}$.

\item [O~I]~6300/H$\alpha$~: knot A2 shows an initial increase in this line ratio, from $\approx 0.2$ for $v_r=-5$~km~s$^{-1}$ to $\approx 0.4$ for $v_r=110$~km~s$^{-1}$.  The line ratio has a $\approx 0.3$ value in the  $v_r=172$~km~s$^{-1}$ velocity channel and then it increases for higher radial velocities, reaching a value of $\approx 0.5$ for $v_r=296$~km~s$^{-1}$.  Knot A1 shows high values of $\approx 0.7$ at $v_r=48$~km~s$^{-1}$, decreasing to values of $\approx 0.1$ at $v_r=296$~km~s$^{-1}$.  The highest values in this line ratio are found to the East of knot A1 in the $v_r=110$~km~s$^{-1}$ channel map.

\item [S~II]~(6716+6731)/H$\alpha$~: this line ratio shows a general trend of decreasing values with increasing radial velocity for $v_r=-5\to 172$~km~s$^{-1}$, and then has similar values for all higher radial velocity channel maps.  Knot A2 has line ratios of $\approx 1$ for $v_r=-5$~km~s$^{-1}$, dropping to $\approx 0.4$ for $v_r=110$~km~s$^{-1}$, and then keeping this value for the higher radial velocity channels.  Knot A1 has a more drastic drop in the line ratio, with values of $\approx 2$ at low velocities, decreasing to $\approx 0.2$ for $v_r=234$~km~s$^{-1}$.

\item [S~II]~6716/6731~: for both A1 and A2, this line ratio shows values of $\approx 0.7$ (corresponding to $n_e\approx 230$~cm$^{-3}$) in the $v_r=-5$~km~s$^{-1}$ map. The line ratio then decreases with increasing radial velocities, with values of $\approx 0.6$ ($n_e\approx 420$~cm$^{-3}$) for $v_r=48$~km~s$^{-1}$ and $\approx 0.55$ ($n_e\approx 650$~cm$^{-3}$) for $v_r=110$~km~s$^{-1}$.  The line ratio at the positions of knots A1 and A2 remains close to the high density limit for all of the radial velocity channel maps with $v_r> 110$~km~s$^{-1}$.  It is clear in the lower heliocentric radial velocity channel maps (with $v_r\leq 110$~km~s$^{-1}$) that knots A1 and A2 are surrounded by a diffuse emitting region (filling most of the observed field in the $v_r=-5$~km~s$^{-1}$ map) with line ratios of $\approx 1$ ($n_e\approx 150$~cm$^{-3}$).

\item [OIII] 5007/H$\alpha$: at the position of condensation A1, this emission line ratio grows continually with radial velocities from -5 km s$^{-1}$ up to +300 km s$^{-1}$.  At the condensation A2, this line ratio is roughly constant for radial velocities from +110 to +238 km s$^{-1}$ and increases for larger radial velocities. This line ratio is clearly larger at the position of condensation A1 than at the knot A2.

\end{itemize}

These results show that the line ratios of HH~32 have strong spatial variations at arcsecond scales, as well as variations as a function of radial velocity.  The ratio maps for the [S~II] lines show that the brighter A1 and A2 condensates are enclosed within a diffuse emitting region with lower electron density than the knots.   This diffuse low electron density region surrounding knots A1 and A2 has low H$\alpha$ emission; it has significant values for both the high excitation [N~II]~6583/H$\alpha$ and the low excitation [O~I]~6300/H$\alpha$ line ratios (see Figure 5).  Indeed, the peak [N~II]~6583/H$\alpha$ and [O~I]~6300/H$\alpha$ values are found in this diffuse region rather than centered on the positions of the A1 or A2 condensations.

Although the behavior of the line ratios is very complex, some of the observed trends are consistent with a simple, plane shock wave interpretation.  For example, the [O~III]/H$\alpha$ ratio in knot A1 grows monotonically with radial velocity, consistent with the fact that this line ratio increases with increasing shock velocities.  Also, the [S~II]~(6716+6731)/H$\alpha$ ratio in the A1 condensation first decreases and then stabilizes as a function of increasing $v_r$, which qualitatively follows the line ratio vs. shock velocity trend predicted from plane-parallel shock models (see Hartigan et al. 1987). However, different regions of HH~32 show different trends in the line ratio versus radial velocity which cannot be easily interpreted with the present plane shock models.   As an example, knot A1 shows a decrease in the [O~I]~6300/H$\alpha$ ratio as a function of increasing radial velocity, while knot A2 has the opposite trend.  Interestingly, for this oxygen line ratio the plane shock models predict a decrease in intensity as a function of shock velocities (from 20 to 90 km/s), followed by a strong increase for larger velocities (from 100 to 300 km/s, see Hartigan et al. 1987).  The observed line ratios of [O~I]~6300/H$\alpha$ in the two knots are not consistent with the present models.

The behavior as a function of $v_r$ of the line emission of knots A1 and A2 can be seen in a more quantitative way in Figure~8.  In this figure, we have plotted the ratios between the H$\alpha$, [O~III]~5007, [O~I]~6300, [N~II]~6583 and [S~II]~(6716+31) line emission of A1 and A2 as a function of $v_r$.  From this graph we see that knot A1 is brighter than A2 for heliocentric radial velocities $v_r<150$~km~s$^{-1}$ in the [O~I]~6300 and [S~II]~(6716+31) lines, and A2 is brighter than A1 for $150<v_r<350$~km~s$^{-1}$.  For $v_r>350$~km~s$^{-1}$, knot A1 is again brighter than A2.

The relative intensities between A1 and A2 in the H$\alpha$ and [N~II]~6583 lines show a similar behavior to [O~I] and [S~II], but have a much shallower dependence on radial velocity for $v_r<300$~km~s$^{-1}$.  The H$\alpha$ fluxes of A1 and A2 differ by at most 20~\%\ in this radial velocity range.  In the H$\alpha$ and [N~II] lines, condensation A1 brightens quite steeply relative to A2 for $v_r>300$~km~s$^{-1}$, becoming twice as bright for $v_r=390$~km~s$^{-1}$.  The relative [O~III] intensity between A1 and A2 shows a different behavior than the other emission lines plotted (the [O~III] channel maps are shown in Figure 7).  The [O~III] emission seen in A1 brightens quite steeply relative to A2 for $v_r$ from 0 to +120 km s$^{-1}$.  The relative intensities remain more or less constant with a mean value of 1.75 for radial velocities in the range from +120 to +300 km s$^{-1}$, and decrease again for larger velocities ($v_r$ $>$ 300 km s$^{-1}$).

\section{A 3/2-D bow shock model}

In order to interpret the observations of HH~32, we consider a traditional ``3/2-D'' bow shock model (Hartigan et al. 1987).   It is assumed that the shock velocity is equal to the component of the pre-bow shock flow velocity normal to the shock surface.  Such models have been used to interpret line profiles of several HH objects (see, e.~g., Hartigan et al. 1987), with 1D or 2D spatial resolution (Raga \& B\"ohm 1985; 1986; Solf et al. 1991; Morse et al. 1992).    To date, one of the most successful applications of the 3/2-D bow shock models comes from the line profile fits to the HH~32 outflow (Solf et al. 1986; Hartigan et al. 1987). 

We consider a bow shock with a functional form:
\begin{equation}
{z\over a}=\left({r\over a}\right)^p\,,
\label{zr}
\end{equation}
where $z$ is measured along the symmetry axis and $r$ is the cylindrical radius.  The constants $a$ and $p$ are the free parameters of the model, with $p$ determining the shape of the bow shock, and $a$ its physical size.  The bow shock is assumed to be moving at a velocity $v_{bs}$ with respect to a uniform pre-shock medium, directed at an angle $\phi$ with respect to the plane of the sky.  Without any loss of generality, we can set $a=1$, and scale the intensity maps predicted from the model in an arbitrary way to produce spatial scales comparable to the ones of HH~32A.

In order for the model to fit closely with the observations, we have assumed that the wings of the bow shock (with the shape given by equation \ref{zr}) suddenly end at a distance $z_{max}$, as measured along the symmetry axis from the head of the bow shock.  We then use the geometry of the system to calculate: 1) the shock velocity, which is the velocity component of the bow shock normal to the surface of the shock, 2) the velocity of the post-bow shock flow which is approximately equal to the normal velocity for the compressions found in radiative shocks, and 3) the intensity per unit area of the bow shock derived by interpolating between the line emissions predicted from a series of plane-parallel shock models with differing shock velocities.  The necessary geometrical construction we have used is described by Raga \& B\"ohm (1986) and in more detail by Hartigan et al. (1987).

We computed the emission using the tabulation of plane-parallel models with a ``self consistent'' pre-ionization and pre-shock density of $n_{pre}=100$~cm$^{-3}$ as described by Hartigan et al. (1987).  In the models, we fixed the bow shock velocity to a $v_{bs}=350$~km~s$^{-1}$ value necessary to obtain line widths comparable to the observed ones, and we adopt an orientation angle of $\phi=70^\circ$ moving away from the observer, as derived by the observed line profiles, radial velocities and proper motions (Raga et al. 1986; Hartigan et al. 1987; Curiel et al. 1997).  We further assume that the pre-shock medium is at rest with respect to the outflow source.

At this point, the power law index $p$ is the only free parameter in the model.  If the shock wings extend to large distances from the head of the bow shock, then we can set $z_{max} \to \infty$ (see equation \ref{zr}).  In this case, the best fits for the model are obtained for $p\sim 2-3$.   However, if we allow a finite value for $z_{max}$, the models more closely resemble the observations.  In particular, we choose a model with $p=2$ and $z_{max}=1.5a$.  From this bow shock model, we have computed velocity channel maps which are used to compare with the observations of HH~32 (shown in Figure 9).  

From comparison of the velocity channel maps shown in Figures 4 and 9, we find that the shock model predicts structures that are in qualitative agreement with the HH~32 observations.  At low radial velocities ($v_r=15\to 90$~km~s$^{-1}$), the model shows strong H$\alpha$ emission distributed in a single arc-like filament.  The observations of HH~32 also show strong H$\alpha$ emission but it is distributed in two or three arc-like structures.  At $v_r=100\to 200$~km~s$^{-1}$, the bow shock model shows fainter H$\alpha$ emission with a circular structure, which becomes more compact for increasing radial velocities.  In this velocity range, the HH~32 data also shows fainter, more compact emission at increasing radial velocities, but with a more complex morphology of knots (see Figure 4).  For $v_r>200$~km~s$^{-1}$, both the model and HH~32 data show an increase in H$\alpha$ emission, which is then followed by an intensity decrease, with tighter knot-like structures at the higher radial velocities. There is a lack of quantitative agreement because the high velocity H$\alpha$ peak occurs at $v_r=308$~km~s$^{-1}$ in the model, but at $v_r=227$~km~s$^{-1}$ in HH~32.  Also, while the model shows a single condensation for $200<v_r<340$~km~s$^{-1}$, HH~32 shows two or three compact condensations, and it does not converge to a single condensation until $v_r\geq 350$~km~s$^{-1}$.  Our simple model considers only one bow shock observed at an angle of $\approx 70^\circ$, comparison with the observations shows that a more complicated structure of 2 (or more) bow shocks would better fit the observations (see Raga et al. 2003).

We computed line ratio maps from our bow shock model and did not obtain results that qualitatively agree with the observations presented in Figures 5, 6 and 7.  This discrepancy is not surprising because of the differences seen in the H$\alpha$ channel maps predicted by the model and observed in HH~32.  The significant structure observed in the line ratio maps and the lack of consistency with our calculations is difficult to interpret uniquely, as a superposition of several shocks along the viewing geometry of HH 32 will complicate the models.

There are two main criticisms for the 3/2-D bow shock model we have utilized to describe the observations of HH 32.   First, it is obvious that the simple, parametrized shape (equation 1) which we have assumed for the emitting shock structure is not appropriate for describing the more complex line of sight superposition of shocks in HH~32.   Secondly, a 3/2-D shock model is based on the assumption that the post-bow shock cooling distance is small compared to the size of the bow shock; this is probably not the case in this scenario.  As can be seen from the models of Hartigan et al. (1987), the cooling distance behind the head of a 250 km/s shock has a value $d_c\sim 10^{16}$~cm, which is comparable to the sizes of condensations A1 and A2.  The model we have used is qualitatively successful at describing the observations of HH~32, but the assumptions on which the 3/2-D bow shock models are based are not ideal for describing this object.  We present and discuss a more detailed 3D simulation of multiple working surfaces in a different paper (Raga et al. 2003)

\section{Conclusions}

We have obtained a high resolution spectrum of HH~32A with 2D spatial resolution which provides a wealth of information about the excitation and kinematics of the HH~32 outflow.  From these spectra, we have constructed channel maps at velocities ranging from $-20$ to $+400$~km~s$^{-1}$ for the H$\alpha$, [O~III]~5007, [O~I]~6300, [N~II]~6583, [S~II]~6716 and [S~II]~6731 emission lines;  the H$\alpha$ maps are presented in Figure 4 and [O~III] is shown in the central panels of Figure 7.  The channel maps of the emission lines of H$\alpha$, [O~I], [N~II], and [S~II] have a qualitatively similar behavior.  In the lower heliocentric radial velocity maps ($v_r<100$~km~s$^{-1}$), HH~32A has a morphology of two or three superimposed arc-like features.  At higher radial velocities, the emission becomes progressively more concentrated, showing two or three compact condensations. Finally, for $v_r>350$~km~s$^{-1}$, a single condensation is seen. 

Figures 5, 6 and 7 show complex structures in the line ratio maps.  The local maxima and minima do not coincide necessarily with the observed knots in the H$\alpha$ images, and the ratios vary for each emission line over the velocity width of the feature.  The complex morphological structures suggest that we are likely observing the emission from several compact knots, superimposed on a diffuse emission component, with the knots and the diffuse component having different spectral properties. This interpretation has been previously suggested to explain 2D spatial resolution spectra of HH~2 (B\"ohm \& Solf 1992).

The observations of HH~32 can be interpreted in a qualitative way with a simple ``3/2-D'' bow shock model.  We constructed model H$\alpha$ velocity channel maps that show a transition between a single arc-like feature at low radial velocities, to a more concentrated and centrally peaked condensation at high radial velocities (Figure 9).  This basic change in morphology is also seen in the velocity channel maps of HH~32A (Figure 4), but the structures in the data are much more complex. 

From comparison with the simple bow shock model, we conclude that HH~32A is likely a superposition of two or three bow shocks, corresponding to different working surfaces along the HH~32 outflow.   HH~32A shows at least two condensations (A1 and A2) with spatial and kinematic properties which resemble the predictions from a single bow shock model.  The fact that we see such a superposition is not surprising; the $\phi\approx 70^\circ$ orientation of HH~32 with respect to the plane of the sky will lead naturally to a line of sight overlap of features along the outflow axis.   HH~32 might be intrinsically similar to the collimated outflow observed in HH~34, we might be viewing two working surfaces catching up with each other as seen in HH~34S (Reipurth et al. 2002; Raga et al. 2002).  The arc-like structures and multiple condensations observed in the data of HH~32A could then correspond to a line of sight superposition of the working surfaces.  We present a study of this scenario based on more detailed 3D numerical simulations in Raga et al. (2003).

\acknowledgments  We thank Inger J{\o}rgensen, GMOS-N instrument scientist, for approving this project for System Verification observations.  A. Riera acknowledges the ICN-UNAM for support during her sabbatical.   A. Riera was supported by the MCyT grant AYA2002-00205 (Spain).  The work of A. C. Raga was supported by the CONACyT grant 36572-E.

\begin{figure}
\epsscale{0.7}
\plotone{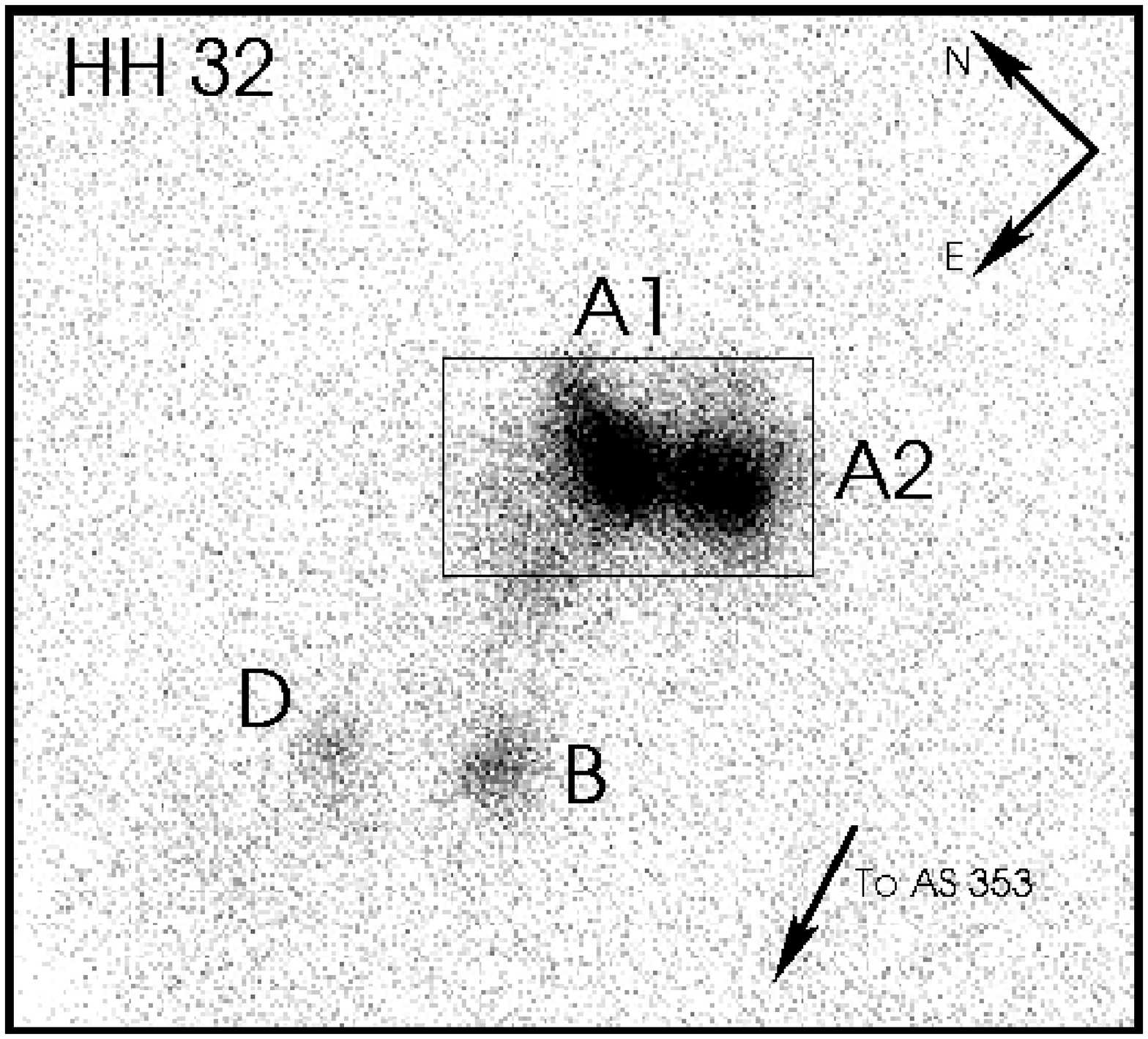}
\caption{A 30 second r-band exposure that was obtained for acquisition of the HH 32 A condensations on the IFU field.  These observations were made at a position angle 155.7 degrees East of North, as indicated by the arrows in the upper right of the figure.  From the notation of Curiel et al. (1997), we have labeled the A1, A2, B and D condensations detected in this short exposure image.  Overplotted around the A1 and A2 condensations is the spatial field of view of the final IFU mosaiced datacube.  We have included an arrow showing the direction to the outflow exciting source, AS 353.
\label{fig1}}
\end{figure}
\clearpage

\begin{figure}
\epsscale{0.7}
\plotone{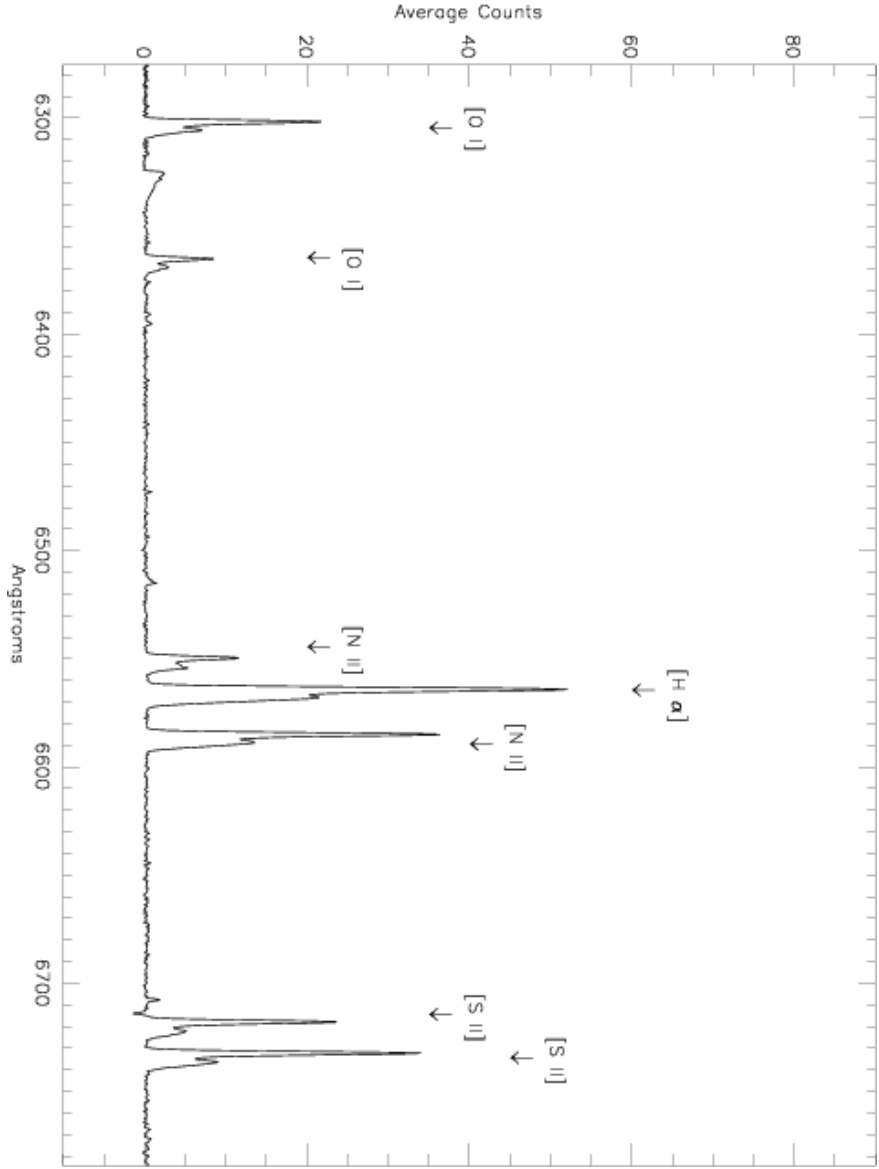}
\caption{A spectrum of the HH 32 A2 condensation obtained by averaging the spectra in the final datacube at the position of this emission knot over the inner 1.$''$5 in the spatial dimension.  The [O~I]~6300/63, [N~II]~6548/83, H$\alpha$ and [S~II]~6716/31 lines are labeled.  The emission features all show a distinct double-peaked structure, consistent with past spectral observations of the condensations.  The "feature" at 6325-6340 A is an instrumental effect that arises from interpolating over the gap between two of the GMOS CCD chips.  The other weak "emission lines" arise from residual averages of imperfectly corrected cosmic rays.  
\label{fig2}}
\end{figure}

\clearpage
\begin{figure}
\plotone{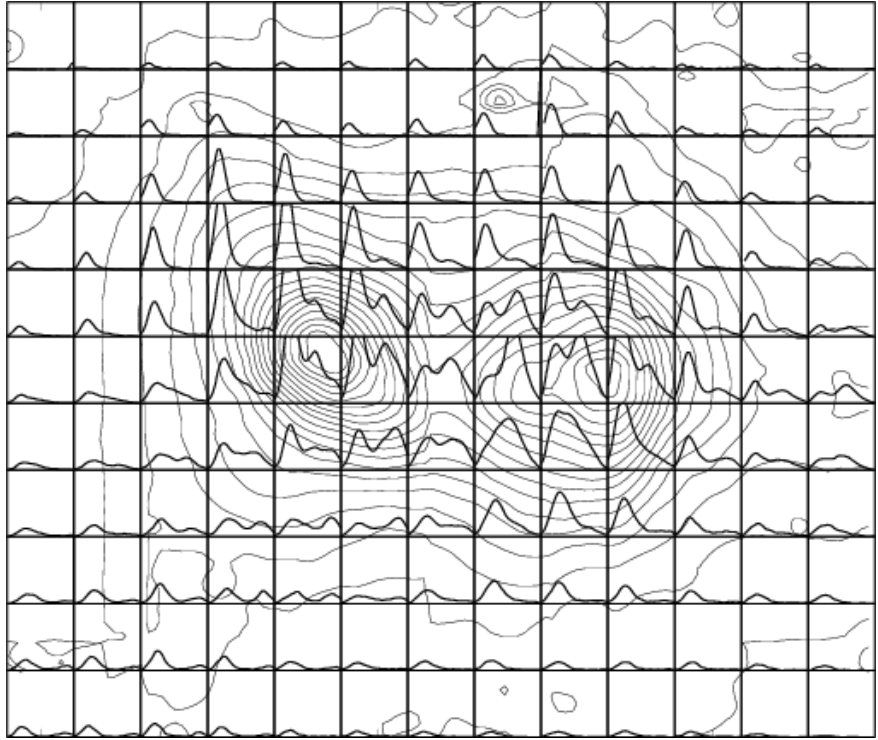}
\caption{The H$\alpha$ image made by integrating the H$\alpha$ emission over wavelength is shown as a linear contour diagram. The line profiles obtained by spatially integrating the emission over boxes of $10\times 10$ pixels (corresponding to $0.5\times 0.5$ arcsec$^{2}$) are plotted as relative intensity (counts) vs. heliocentric radial velocity (a radial velocity from $-$20 to 430 km s$^{-1}$ is shown).  The spatial field presented in this figure is $6\farcs5$ $\times$ $5\farcs5$, the left side of the figure was cropped in the x direction because this region has low H$\alpha$ flux levels.
\label{fig3}}
\end{figure}
\clearpage

\begin{figure}
\plotone{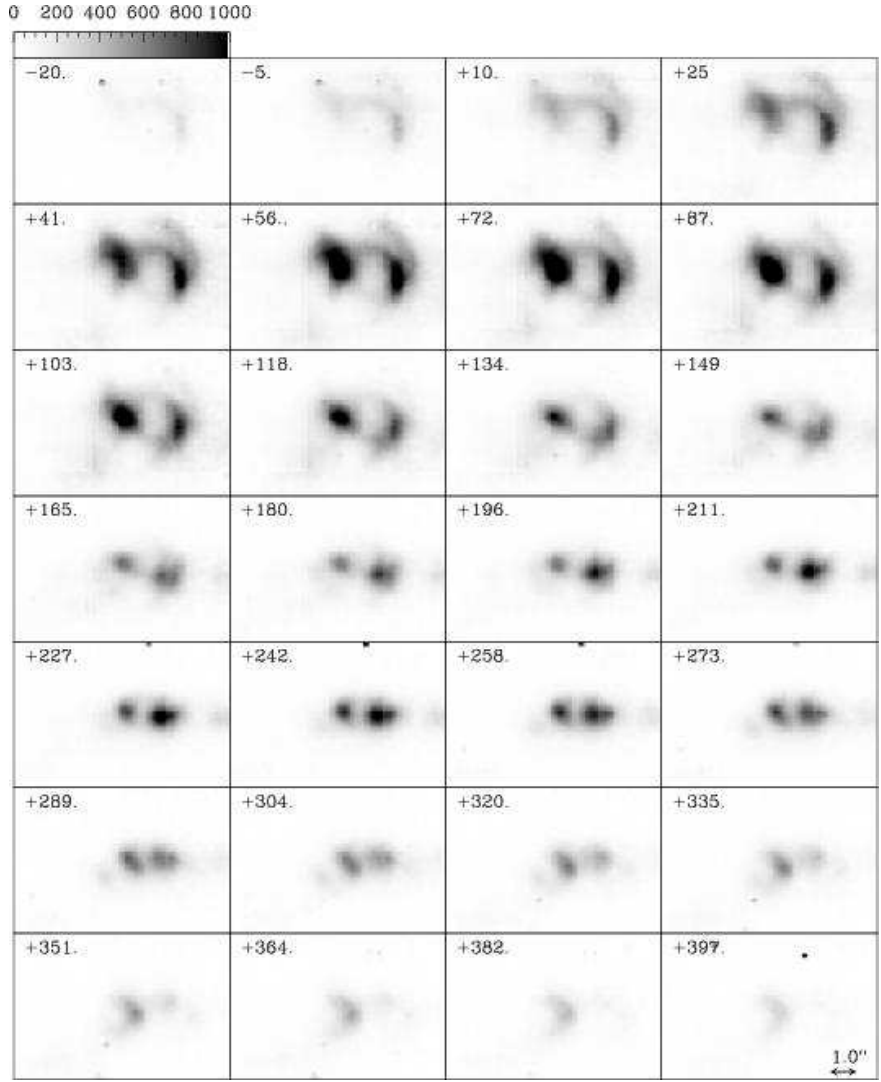}
\caption{The observed H$\alpha$ radial velocity channel maps are plotted with a linear greyscale. The heliocentric radial velocities range from $-$20 to $+$397 km s$^{-1}$. The central heliocentric velocity of each channel is shown in the corresponding panel.  The spatial scale is shown in the $+$397 km s$^{-1}$ panel.  The intensity scale displayed at the top left of the figure is in counts.
\label{fig4}}
\end{figure}
\clearpage

\begin{figure}
\caption{Left panels: H$\alpha$ channel maps centered at heliocentric radial velocities of $-$5, $+$48, $+$110, $+$172, $+$234, $+$296, and $+$358 km s$^{-1}$ and with velocity widths of  47 km s$^{-1}$ (see section 4). Central panels: [N II] 6583/H$\alpha$ ratio maps obtained from the channel maps at the heliocentric velocities mentioned above.  Right panels: [O I] 6300/H$\alpha$ ratio maps obtained from the channel maps at the heliocentric velocities mentioned above. The color scales for the the plots are given by the top bars.  The line ratio maps were computed with intensity maps with a $5\times 5$ pixel binning.  The lower threshold for each emission line that was used in plotting the flux ratios was set at 300 counts to limit noise in the maps from the low flux regions.
\label{fig5}}
\end{figure}
\clearpage

\begin{figure}
\caption{The same as Figure 4 but for the [S II] 6716/6731 (central panels) and [S II](6716+6731)/H$\alpha$ line ratios (right panels).  The color scales for the the plots are given by the top bars for the left two columns and by the bars at the right for the [S~II]/H$\alpha$ ratio.  The lower threshold for each emission line used in plotting the flux ratios was set at 300 count to limit noise in the maps from the low flux regions.
\label{fig6}}
\end{figure}
\clearpage

\begin{figure}
\caption{Left panels: H$\alpha$ channel maps (as in Fig. 4).  Central panels: [OIII] 5007 \AA\ channel maps centered at heliocentric radial velocities of -5, +48, +110, +172, +234, +296, and +358 km s$^{-1}$. Right panels: [O III]/H$\alpha$ ratio maps obtained at the heliocentric velocities mentioned above, with the logarithmic color scale given by the top bar. The line ratio maps were computed using the intensity maps with a 5 x 5 pixel binning.  The lower threshold for the [O~III] emission line for plotting the flux ratios was set at 1000 counts to limit the noise in the maps from the lower flux regions.
\label{fig7}}
\end{figure}

\begin{figure}
\epsscale{0.60}
\plotone{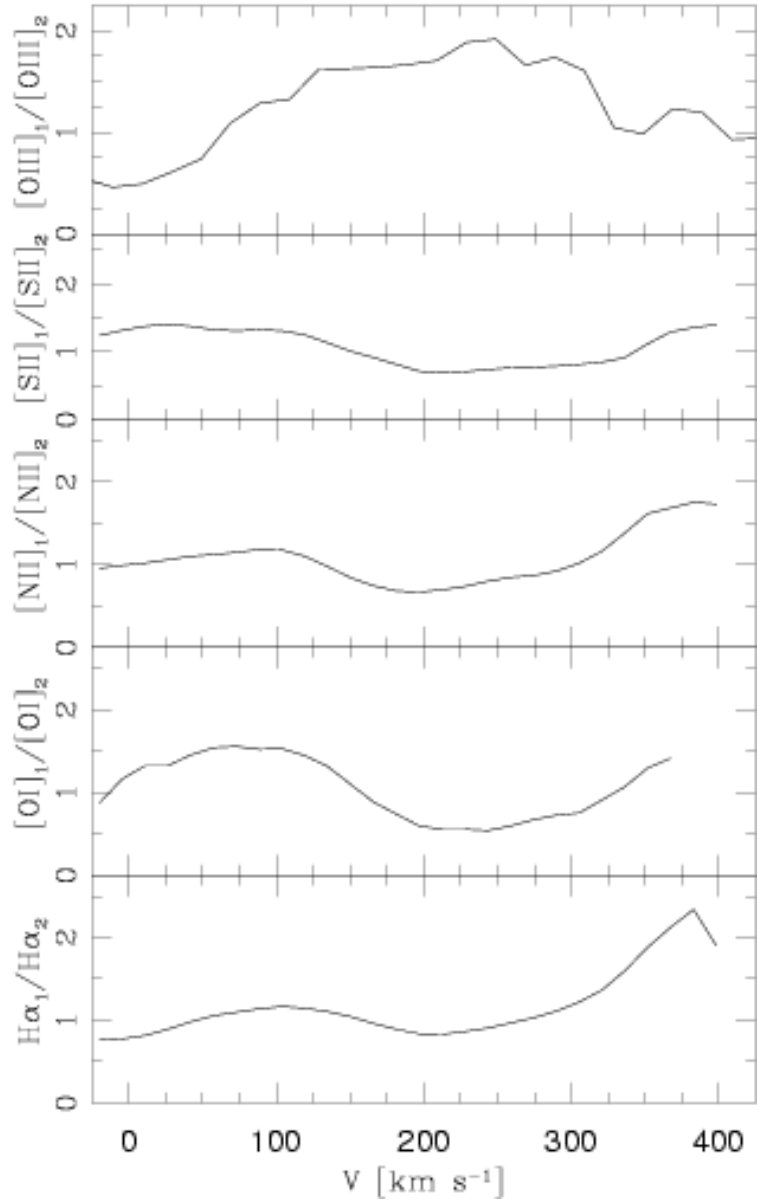}
\caption{Ratios between the emission of condensations A1 and A2 in the H$\alpha$, [O I] 6300, [N II] 6583, [S II] (6716+6731) and [O III] 5007 lines, as a function of the heliocentric radial velocity.  
\label{fig8}}
\end{figure}
\clearpage

\begin{figure}
\epsscale{0.9}
\plotone{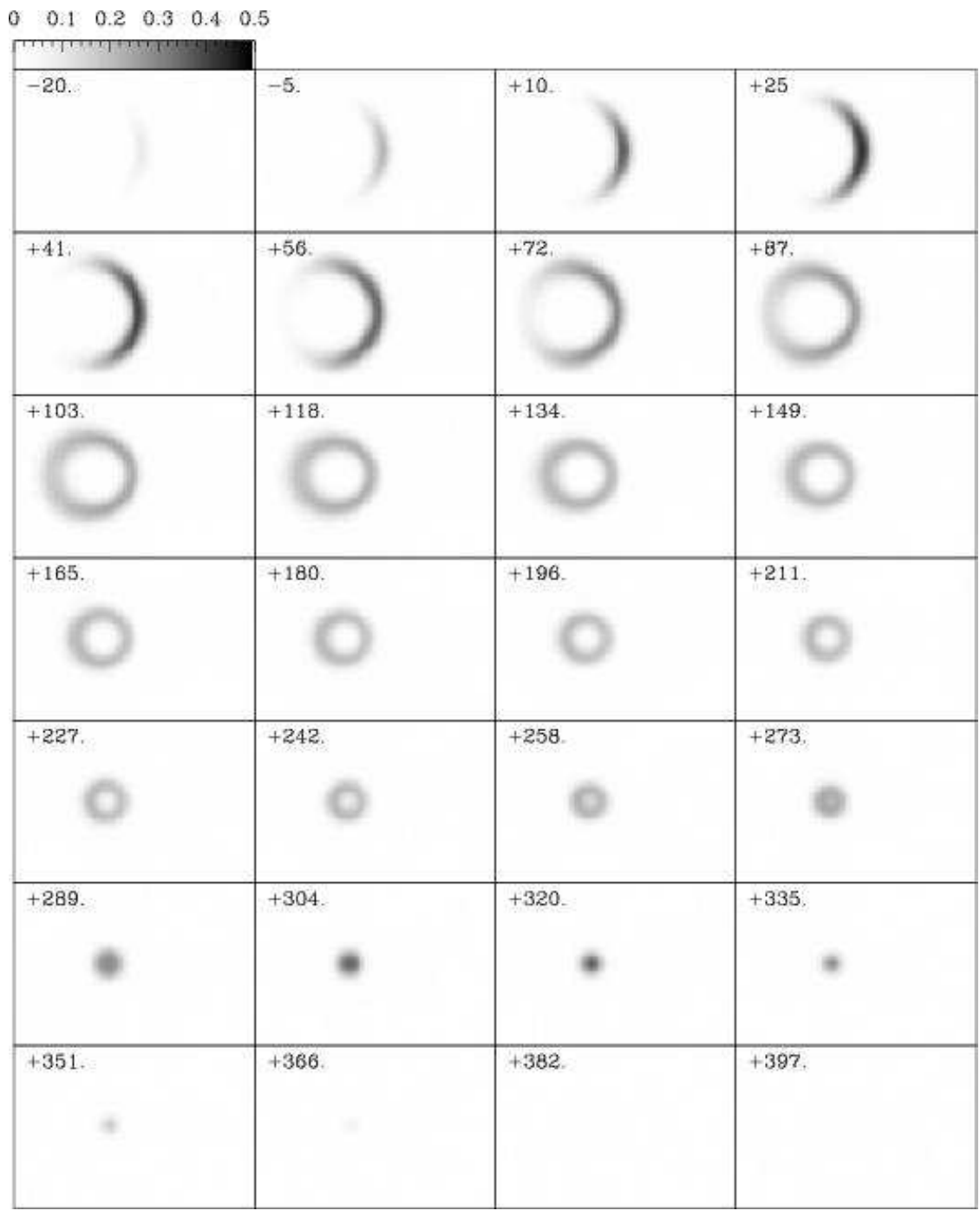}
\caption{H$\alpha$ velocity channel maps computed from the 3/2-D bow shock model described in the text. The intensities have been normalized so that the peak intensity in the $v_r=+29$~km~s$^{-1}$ map has a value of 1. We have then convolved the normalized maps with a 2D Gaussian in order to simulate a seeing of FWHM$=0''.5$.  On a qualitative level, the model agrees fairly well with the data presented in Figure 4 - the intensity has a double peaked structure, with one peak at $v_r=40-70$~km~s$^{-1}$ and one at $v_r=270-320$~km~s$^{-1}$.  The model also shows an arc-like structure at lower radial velocities, and more concentrated knots of emission at higher $v_r=$.
\label{fig9}}
\end{figure}

\end{document}